\newcommand{\scr}[1]{\scriptsize{#1}}

\documentclass{aa}
\usepackage{url}
\usepackage{hangcaption}
\usepackage{natbib}
\usepackage{graphicx}

\begin{document}

\title{Prospects for accurate distance measurements of pulsars with the SKA:
Enabling fundamental physics}
\author{R.~Smits \inst{1,2} \and S.J.~Tingay \inst{3} \and N.~Wex \inst{4}
  \and M.~Kramer \inst{4} \and B.~Stappers \inst{1}}
\authorrunning{Smits et al.}
\titlerunning{Prospects for accurate distance measurements of pulsars with the SKA}

\offprints{R. Smits \email{Roy.Smits@\allowbreak
    manchester.\allowbreak ac.uk}}

\institute{Jodrell Bank Centre for Astrophysics, School of Physics and
  Astronomy, University of Manchester, Oxford Road, Manchester M13
  9PL, UK \and ASTRON, Oudehoogeveensedijk 4, Dwingeloo, 7991PD, The
  Netherlands \and International Centre for Radio Astronomy Research -
  Curtin University, 6102 Bentley, WA, Australia \and
  Max-Planck-Institut f\"ur Radioastronomie, Bonn, Germany, Auf dem
  Huegel 69, 53121 Bonn, Germany}

\abstract { Parallax measurements of pulsars allow for accurate
  measurements of the interstellar electron density and contribute to
  accurate tests of general relativity using binary systems. The
  Square Kilometre Array (SKA) will be an ideal instrument for
  measuring the parallax of pulsars, because it has a very high
  sensitivity, as well as baselines extending up to several thousands
  of kilometres. We performed simulations to estimate the number of
  pulsars for which the parallax can be measured with the SKA and the
  distance to which a parallax can be measured. We compare two
  different methods. The first method measures the parallax directly
  by utilising the long baselines of the SKA to form high angular
  resolution images. The second method uses the arrival times of the
  radio signals of pulsars to fit a transformation between time
  coordinates in the terrestrial frame and the comoving pulsar frame
  directly yielding the parallax. We find that with the first method a
  parallax with an accuracy of 20\% or less can be measured up to a
  maximum distance of 13\,kpc, which would include 9\,000 pulsars. By
  timing pulsars with the most stable arrival times for the radio
  emission, parallaxes can be measured for about 3\,600 millisecond
  pulsars up to a distance of 9\,kpc with an accuracy of 20\%.
  }

\authorrunning{Smits et al.} 
\titlerunning{Prospects for accurate distance measurements of pulsars with the SKA}
\maketitle

\section{Introduction}

Distance measurements have always played a hugely important role in
most aspects of astronomy, on both Galactic and extragalactic scales,
but are often very difficult to access observationally. Arguably, the
most reliable measurements are naturally obtained from parallax
measurements when the annual movement of the Earth around the Sun can
be used to detect a variation in the apparent source position. For
nearly all sources and applications, this involves imaging the source
against the background sky. This can be done at optical wavelengths
or, usually more precisely, at radio frequencies using Very Long
Baseline Interferometry (VLBI). In the case of radio pulsars however,
a second type of parallax can be measured using timing measurements of
the radio pulses. Here, distances are retrieved by detecting a
variation in pulse arrival times at different positions of the Earth's
orbit caused by the curvature of the incoming wavefront. In contrast
to an imaging parallax, the precision of timing parallax measurements
is highest for low ecliptic latitudes and lowest for the ecliptic
pole. Pulsar timing of binary pulsars offers further possibilities to
determine the distance from a secular or annual variation in some
orbital parameters. In all cases, however, the methods are usually
limited to relatively nearby sources, but improvement in telescope
baseline lengths and sensitivity promise improvements in precision and
hence the accurate measurement of distances.

The most significant advance will be achieved by the Square Kilometre Array
(SKA), which is being planned as a multi-purpose radio telescope constructed
from many elements, both dishes and aperture arrays, leading to a
total collecting area approaching 1 million square metres
\citep[e.g.][]{Dewdney09}. In the frequency range of 200\,MHz to
3\,GHz, the sensitivity of the SKA will be around 10\,000\,m$^2$/K,
but depending on the design that will be chosen the sensitivity could
be slightly higher or lower \citep[see][]{Schilizzi2007}. The core of
the SKA will be densely filled with 50\% of the collecting area within
a 5\,km diameter area. The remaining elements will be placed at
locations extending up to several thousands of kilometres from the
core. The field of view (FoV) of the dishes will be around
0.64\,deg$^2$ at 1.4\,GHz for a single-pixel receiver and might be
extended up to 20\,deg$^2$ by means of phased array feeds, although
for the long baselines only the single-pixel receivers will be
available. 

Having both a dense core and long baselines allows the SKA to both
find radio pulsars and to perform accurate astrometric measurements on
them. In particular, the SKA will be able to perform parallax
measurements on a great number of pulsars, allowing accurate
determination of their distances from the Earth. Most distances to
pulsars are currently derived from the measured column density of free
electrons between the pulsars and Earth, known as the dispersion
measure (DM). Using a model for the interstellar medium and its free
electron content, the DM can be converted into a distance.  However,
the Galactic electron distribution is not accurately known, so that
the distance derived from DM values is often uncertain, the errors in
the distance estimates being as large as 100\% \citep[see
  e.g.][]{Deller09}.

The SKA will perform a Galactic census of pulsars \citep{ckl+04} that
will potentially yield a population of 20,000 or more radio pulsars
across the Galaxy \citep{Smits09}. Parallax measurements of many of
these pulsars with the SKA will provide direct measurements of their
distances, which will then allow accurate studies of the ionised
component of the interstellar medium, given the measurement of the
DM. In this way, the interstellar electron model can be calibrated and
will not only provide reliable estimates of the DM distances of
pulsars without a parallax measurement in return, but will
simultaneously provide a map of the free electron content in the Milky
Way that can be combined with HI and HII measurements to unravel the
Galactic structure and the distribution of ionised material. Combined
with Faraday rotation measurements of the same pulsars, the Galactic
magnetic field can also be studied in much greater detail as today,
since the precise distance measurements potentially allow us to
pin-point field-reversals occurring with some accuracy.

The determination of accurate pulsar distances is also important for
those pulsars that are part of binary systems, in particular for those
with another compact object, such as the Double pulsar
\citep{Burgay2003,Lyne2004}.  Relativistic effects can be used to
determine the distance to some of these systems when the validity of
general relativity is assumed. In reverse, to perform precision tests
of general relativity, kinematic effects have to be removed for which
it is often required to know the distance precisely
\citep{lk05,ksm+06,Deller09}.

In this paper, we review the possible ways to determine pulsar
distances with the SKA. To compare the technical capabilities with the
expected pulsar population, we simulate an SKA pulsar survey,
following \citet{Smits09}. Using the results of this simulation we
calculate the accuracy with which the SKA can determine the distance
to each pulsar by measuring its parallax. We examine two basic methods
for performing the astrometry, the first that uses imaging and the
long baselines of the SKA, and the second that uses the timing of the
pulsar radio signal. Here we distinguish between parallax measurements
(applicable to all pulsars) and distance-related effects for binary
pulsar parameters.

In Sect. \ref{sec:simulation}, we describe the simulation of the SKA pulsar
survey. The two different methods for measuring the parallax of
pulsars are explained in ref{sec:parallax}, where we also explain
how we estimate the capability of the SKA to measure the parallaxes of
pulsars. The results are presented in \ref{sec:results}. Finally,
we discuss our findings in \ref{sec:discussion}.

\section{Simulation of an SKA pulsar population}
\label{sec:simulation}

To demonstrate the SKA capabilities not only in principle but to also
estimate how many pulsars would offer the highest astrometry
precision, we gauge the overall SKA performance for pulsar astrometry
by simulating the population of pulsars that can be expected to be
involved in the experiments.

The computational requirements for data analysis limit the amount of
collecting area that can be used in an SKA pulsar survey designed to
find new pulsars. According to \citet{Smits09}, at the beginning only
the inner 1-km of the SKA core can be used, which contains $\sim$20\% of
the total collecting area. We performed the same simulation as
performed by \citet{Lorimer2006}, using their Monte Carlo simulation
package. In their study, \citet{Lorimer2006} used the results from
recent surveys with the Parkes multi-beam system to derive an
underlying population of 30\,000 normal pulsars with an optimal set of
probability density functions (PDFs) for pulsar period ($P$), 1400-MHz
radio luminosity ($L$), Galactocentric radius ($R$), and height above
the Galactic plane ($z$). For our simulations, we adopt the PDFs of
model C$^\prime$ in \citet{Lorimer2006}. The intrinsic pulse width of
each pulsar follows the power law relationship with spin period given
in Eq. (5) of~\citet{Lorimer2006} (see also their Fig. 4). To
compute the expected DM and scatter broadening effects on each pulse,
we use the NE2001 electron density model. To scale the
scatter-broadening time to arbitrary SKA survey frequencies, we adopt
a frequency power-law index of $-4$~\citep{lkm+01}.

We then simulated an all-sky pulsar survey at 800\,MHz with a
bandwidth of 500\,MHz and 20\% of the sensitivity of the full SKA
(i.e. 2\,000\,m$^2$/K).  We consider a pulsar to be detected by a
given configuration if its flux density exceeds the threshold value,
and its observed pulse width is less than the spin period.

\begin{figure}[htb]
  \includegraphics[width=0.35\textwidth,angle=-90]{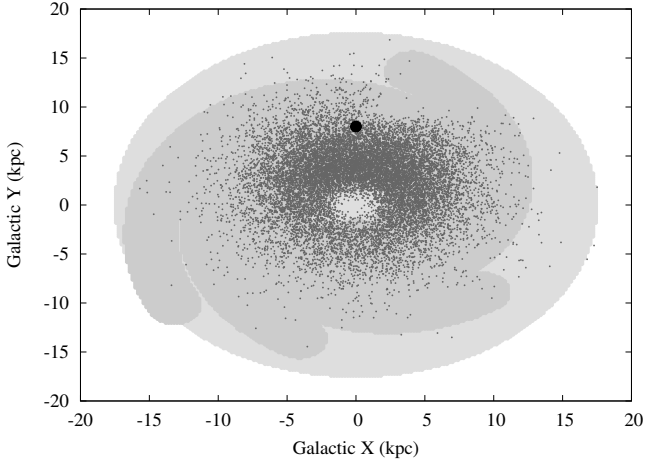}
  \caption{Distribution of the 15\,000 normal pulsars detected in the
    simulation (see text for details). The centre of the plot
    corresponds to the centre of the Galaxy. The grey areas mark the
    inner and outer disks and spiral arms used in the NE2001 electron
    density model. The black circle indicates the position of the
    Earth. }
  \label{fig:population}
\end{figure}

Since the core of the SKA will be located near the latitude of
--30$^\circ$, we limited the simulation to only detecting pulsars with a
declination of 50$^\circ$ or less. This led to the detection of
15\,000 normal pulsars in our simulation. In addition, the SKA would
find 6\,000 MSPs. Figure~\ref{fig:population} shows the distribution of
these pulsars in the Galaxy. The void at the centre of the figure is
due to the extreme scattering of the radio signal at the Galactic
centre, which decreases the sensitivity of detection. The degree of
scattering near Sgr A* is not yet known but we can expect that
frequencies above 10 GHz may be required to detect pulsars very close
to the central supermassive black hole \citep{ckl+04}.

\section{Distance measurements}
\label{sec:parallax}

The parallax and hence distance of a stellar object is classically
obtained by measuring the celestial coordinates of the object on the
sky at different orbital positions of the Earth. Here we examine two
methods for measuring the parallax of pulsars with the SKA, imaging
parallax and timing parallax, before we also briefly discuss the
``orbital'' parallax measurements of binary pulsars.

\subsection{Imaging observations}
\label{sec:imaging}

The straightforward method for measuring the parallax of pulsars is to
measure the position of the pulsar on the sky over time by means of
imaging.  \citet{fom04} estimate the astrometric capabilities of the
SKA by means of interferometry using the stated design goals of this
proposed instrument \citep{Schilizzi2007}.  They outline some of the
calibration techniques for astrometry that could be employed by the
SKA and discuss the effects of limited dynamic range on astrometry;
they conclude that with multiple calibrator sources available at
separations of a few arcminutes the astrometric accuracy that the SKA
can potentially obtain is $\theta_{a}\sim\theta/1000$, where
$\theta_{a}$ denotes the astrometric accuracy, $\theta$ is the imaging
resolution obtained by the array (typically FWHM of the synthesised
beam) and the factor 1\,000 is the dynamic range. At the frequency of
1.4 GHz and a 3\,000\,km maximum SKA baseline, the potential
astrometric accuracy is approximately 15\,$\mu$as. However, for many
pulsars the limiting factor will be given by the limited SNR of the
pulsar detection. We therefore estimate the astrometric accuracy as
the maximum of $\theta/1000$ and $\theta/\mathrm{SNR_{pulsar}}$, where
$\mathrm{SNR_{pulsar}}$ is the SNR of the pulsed flux from the pulsar
over an integration time of 12 hours. To achieve this SNR requires the
correlator to perform de-dispersion and gating over the on-pulse of
the pulsar. Therefore, an ephemeris of the pulsar is required that
can be obtained from regular timing observations.

As of the end of 2009, parallaxes for approximately 30 pulsars have
been obtained using VLBI \citep[for a summary see][]{vlm10}.
Typically about six measurements over a period of two years are
required to measure the parallax and proper motion of a single pulsar,
typically requiring a total observing time of 72 hours of
observational time per pulsar (12 hours per VLBI
observation). Measuring the parallax of multiple pulsars in the FoV of
the telescope involves using the correlator to perform de-dispersion
and gating for every pulsar in the FoV, which is possible for current
VLBI observations using the capabilities of new software correlation
systems.

\subsection{Timing observations}

A second set of methods for performing astrometry of a pulsar involves
the accurate timing of the pulsar radio signals, i.e.~the measurement
of the pulse times-of-arrivals (TOAs) at the telescope. Here we can
distinguish between three methods, all relying on very precise TOA
determination: {\em a)} parallax measurements using the Earth orbit,
{\em b)} parallax measurements for binary pulsars using the Earth
orbit and that of the pulsar, and {\em c)} distance estimates of
binary pulsars based on the comparison of observed orbital parameters
with those predicted by general relativity.

\begin{figure*}[htb]
\centering
\begin{tabular}{cc}
\includegraphics[width=0.55\textwidth]{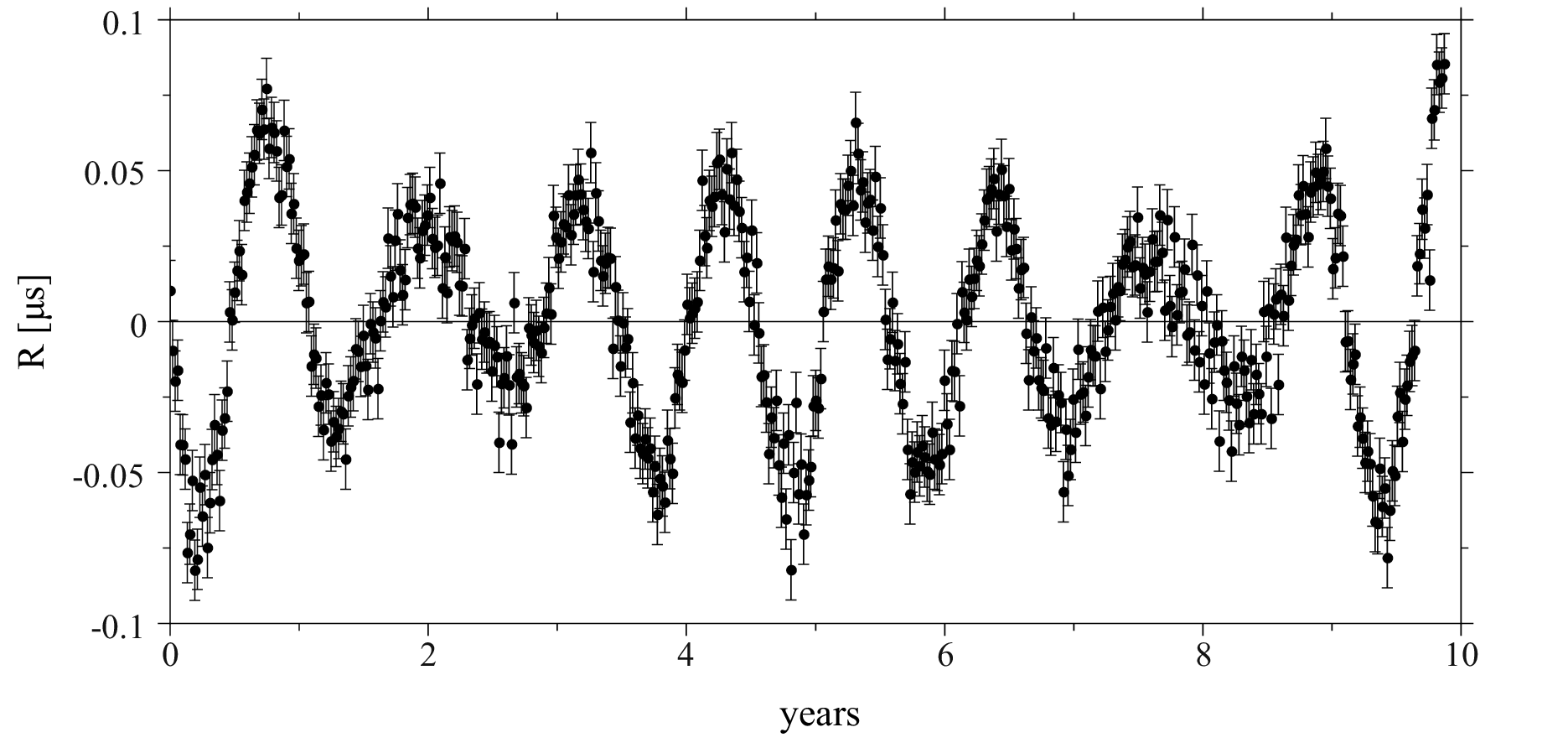} &  
\includegraphics[width=0.4\textwidth]{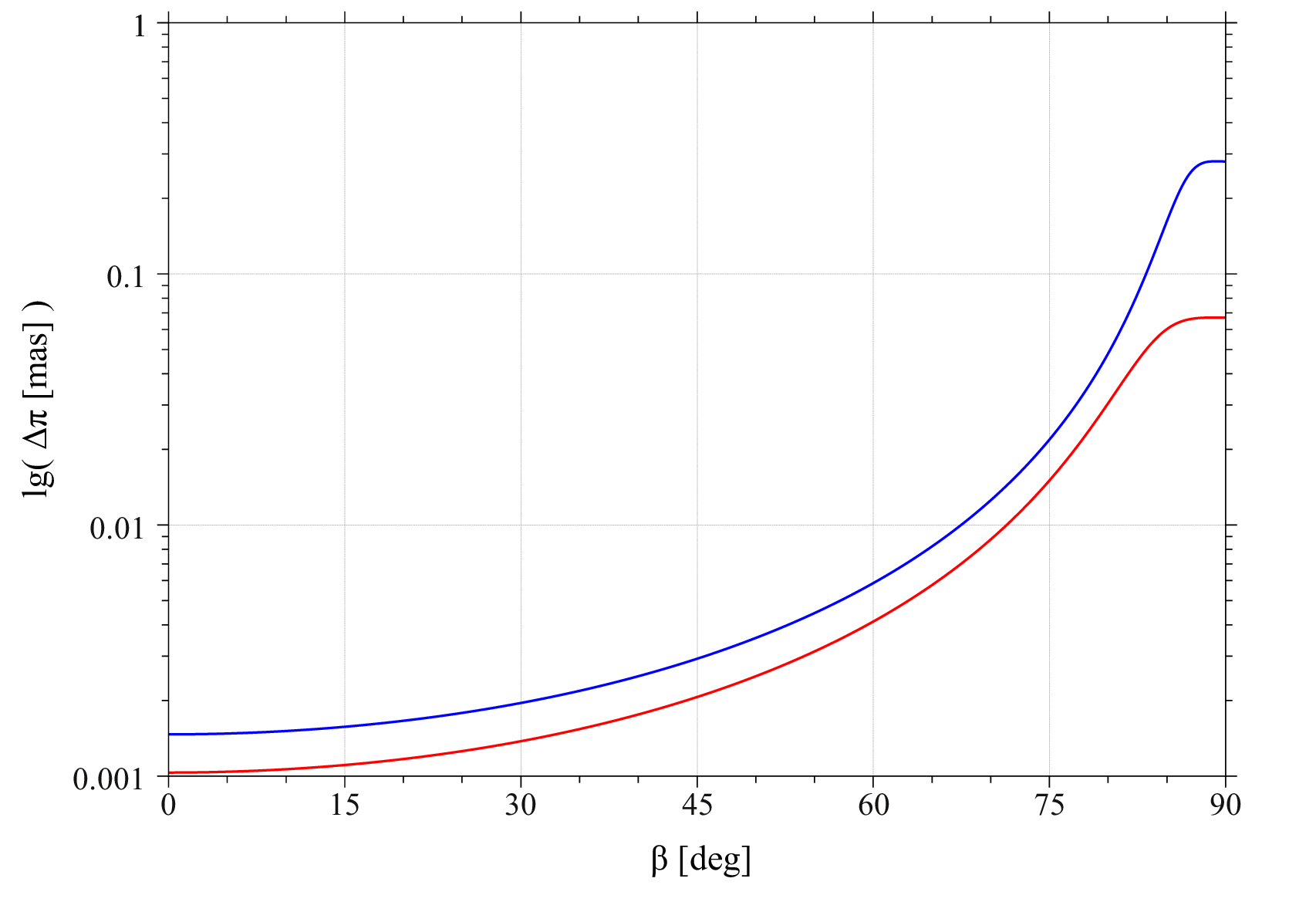}
\end{tabular}

\caption{{\em Left)} Post-fit residuals for a pulsar at a distance of
  200\,pc located at the pole of the ecliptic after fitting for
  position and proper motion but not for parallax. Remaining structure
  in the TOAs is clearly visible.  {\em Right)} Error in the timing
  parallax as a function of the ecliptic latitude $\beta$ for weekly
  timing of a pulsar to a TOA with 10\,ns error.  The blue and red
  lines correspond to 5 years and 10 years of timing, respectively.}
\label{fig:wex1a}
\label{fig:wex1c}
\end{figure*}

\subsubsection{Timing parallax}

The ``classical'' timing parallax measurement utilises the fact that
the wave front curvature of a pulsar signal is directly related to the
distance of the source. For an infinitely distant source, the wave
front is planar and, after taking into account Earth's orbit, the TOA
would essentially be the same at all orbital phases of the Earth. For
a finite distance, however, the curvature of the wavefront introduces
an annual periodic change in the apparent direction, hence a
six-monthly periodicity in the TOAs due to the changing path-length
from the pulsar to the Earth \citep{ls04,lk05}. The apparent change in
direction is more easily measurable for low ecliptic latitudes -- in
contrast to an imaging parallax -- as the amplitude of the TOA
variation scales with $\cos^2\beta/d$, where $\beta$ is the ecliptic
latitude and $d$ the distance to the pulsar.

Performing simulations of parallax measurements (see
Fig.\ref{fig:wex1a}) deviations from the simple $\cos^2\beta$ scaling
become apparent: it is still possible to measure a parallax with a
finite precision even at high ecliptic latitudes.  As the Earth's
orbit deviates from a circular shape because of a small eccentricity
and deviations caused by other masses, non-linear terms depending on
$e^2$ cause differences in the arrival times that are measurable if
the precision is sufficiently high.

Figure~\ref{fig:wex1a} shows the absolute uncertainty in a parallax
measurement for an observing span of 5 years and 10 years,
respectively. In both cases, weekly TOAs with 10\,ns precision are
assumed, which is challenging but not impossible with SKA sensitivity
(see Section~\ref{sec:discussion}). In these cases, it will be
possible to determine the parallax of a pulsar with a precision of
$\sim1\mu$as for pulsars near the ecliptic, and still with a precision
of 300\,$\mu$as and 70\,$\mu$as, respectively, at the ecliptic
pole. In other words, in principle the SKA should allow us to achieve
a distance measurement with an uncertainty of less than 5\% out to
about 30 kpc! If the TOA errors are larger, we can still obtain a
precision of 20\% out to 30\,kpc with 50\,ns TOA errors and to 15\,kpc
with 100\,ns TOA errors, respectively. While a timing precision of
100\,ns has been achieved already for some pulsars, in reality the
timing precision will decrease with distance as the pulsars become
fainter. We discuss this further in Sect.~\ref{sec:results}.

\subsubsection{Orbital parallax}

If a pulsar happens to be in a binary system, the pulsar orbit will be
viewed under slightly different angles from different positions of the
Earth's annual orbit. The result is a periodic change in the observed
longitude of periastron, $\omega$, and the 
projected semi-major axis $x=a\sin i$, where $a$ is the semi-major axis
of the pulsar orbit. This effect, known as the {\em annual orbital
  parallax} \citep{kop95,vb03}, causes a variation in $x$ and $\omega$ given by
\begin{eqnarray}
x(t) &=& x_0\left[
1-\frac{\cot i}{d}\; \vec{r}_{\rm SSB}(t)\cdot \vec{J'}
\right]\!\!, \\
\omega(t) & = & \omega_0 - \frac{\csc i}{d} \vec{r}_{\rm SSB} \cdot
\vec{I'} \!\!,
\end{eqnarray}
where $\vec{J'}=-\sin\Omega_{\rm asc} \vec{I_0}+\cos\Omega_{\rm
  asc}\vec{J_0}$ and $\vec{I'}=-\cos\Omega_{\rm asc}
\vec{I_0}+\sin\Omega_{\rm asc}\vec{J_0}$.  The vectors $\vec{I_0}$ and
$\vec{J_0}$ and the longitude of the ascending node $\Omega_{\rm
  asc},$ define the orbital orientation in space \citep[see
  e.g. Figure 8.3 in][]{lk05}.  Since the amplitude of this effect
depends on the distance $d$, a detection of the annual orbital
parallax can also lead to an accurate pulsar distance measurement if
the orientation of the orbit can be determined, e.g.~from aberration
and/or spin-precession effects. Furthermore, a significant proper
motion in the sky can lead to measurable {\em secular} changes in $x$
and $\omega$, which can be used to constrain the orientation of the
pulsar orbit \citep{kop96}. For this, one needs to ensure that the
other contributions to these secular changes are either well
understood or well below the measurement precision of $\dot x$ and
$\dot\omega$. For instance, the secular variation in $x$ can be the
result of various effects that can be summarised as:
\begin{equation}
\label{equ:XDOBS}
    \left(\frac{\dot x}{x} \right)^{\rm obs}\!\!\!\!=
    \left(\frac{\dot x}{x}\right)^{\rm PM}\!\!\!+
    \left(\frac{\dot x}{x}\right)^{\rm GW}\!\!\!+
    \frac{{\rm d}\varepsilon_{\rm A}}{{\rm d}t}-
    \frac{\dot D}{D},
\end{equation}
where the contributions to the observed semi-major axis derivative are a
change in the orbital inclination due to the proper motion, as outlined above,
the emission of gravitational waves (GW), a varying aberration, ${\rm
  d}\varepsilon_{\rm A}/{\rm d}t$, and a changing (radial) velocity
contribution, $-{\dot D}/D$.

The last term of Eq.~(\ref{equ:XDOBS}) is a combined effect of a
changing Doppler shift caused by a relative acceleration in the
gravitational field of the Galaxy or a globular cluster and an
apparent acceleration due to the proper motion across the sky
\citep{dt91}
\begin{equation}
\label{equ:DOPD}
   -\frac{\dot D}{D} =
           \frac{1}{c} \: \vec{K}_{\rm 0} \cdot (\vec{a}_{\rm PSR} -
  \vec{a}_{\rm SSB})
          + \frac{V_{\rm T}^2}{c\: d}\; ,
\end{equation}
where $\vec{K}_{\rm 0}$ is the unit vector from the Earth towards the pulsar,
and $\vec{a}_{\rm PSR}$ and $\vec{a}_{\rm SSB}$ are the Galactic accelerations
at the location of the binary system and the Solar System barycentre.  The
last term including the transverse velocity, $V_{\rm T}$, is known as the
Shklovskii term. Consequently $\dot D/D$ depends on the distance to the
system, and in principle could provide another way to estimate the distance to
the pulsar. In practice, this term is generally below the measurment precision,
and anyway difficult to separate from the other contributions. However, the
acceleration $\dot D/D$ can play an important role in the distance measurment
as outlined in the next section.

\begin{figure}[htb]
\centering
\includegraphics[width=0.5\textwidth]{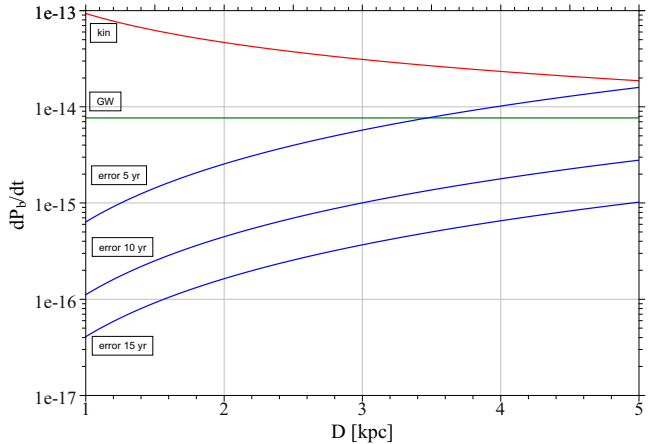} 
\caption{Decrease in orbital period for a pulsar in a 1-d orbit with a
 $0.3M_\odot$ companion and a transverse velocity of 100 km
 s$^{-1}$. The green curve shows the contribution from gravitational
 wave damping, while the blue curve is drawn for different time spans
 of observations, namely 5, 10, and 15 yr, respectively. We assume
 weekly observations, and a 10 ns TOA uncertainty that increases
 quadratically with distance to account for the decreasing flux density.}
\label{fig:wex4a}
\end{figure}

\subsubsection{Using general relativity for distance measurements or
  kinematic parallax}

While the observed $\dot x$ can be contaminated by a variety of
effects (see Eq.~\ref{equ:XDOBS}), the observed orbital decay rate of
an orbit, $\dot{P}_{\rm b}$, is not affected by aberration or proper
motion effects.  If the intrinsic decay rate is determined purely by
the loss of orbital energy due to GW emission, one can predict
$\dot{P}_{\rm b}$ if the pulsar and companion masses are obtained via
pulsar timing thanks to the measurement of {\em post-Keplerian (PK)
  parameters}. If two PK parameters are inferred, for instance from
relativistic orbital precession, gravitational redshift, or a Shapiro
delay, the masses can be determined \citep[see e.g.][]{lk05}) and we
can compute the value of $\dot{P}_{\rm b}$ expected for GW emission
assuming that general relativity (GR) is correct. In this case
\citep{dd85,dd86},
\begin{equation}
\dot{P}_{\rm b}^{\rm GW} = -\frac{192\pi}{5} T_\odot^{5/3}  \left( \frac{P_{\rm b}}{2\pi} \right)^{-5/3} 
               f(e)
               \frac{m_{\rm p}m_{\rm c}}{(m_{\rm p} + m_{\rm c})^{1/3}}, \label{equ:pbdot}
\end{equation}
where all masses are expressed in solar units by using the constant
$T_\odot=(GM_\odot/c^3)=4.925490947$~$\mu$s, $G$ is Newton's
gravitational constant, $c$ the speed of light, and
\begin{equation}
\label{equ:fe}
f(e) = \frac{\left(1+(73/24)e^2+(37/96)e^4\right)}{(1-e^2)^{7/2}}.
\end{equation}
However, the observed value of $\dot{P}_{\rm b}$ will often differ from the
expected intrinsic value (Eq.~\ref{equ:pbdot}) by an additional term
that is again given by the relative acceleration, namely
\begin{equation}
\label{equ:PBDOBS}
    \left(\frac{\dot P_{\rm b}}{P_{\rm b}} \right)^{\rm obs}\!\!\!\!\!=
    \left(\frac{\dot P_{\rm b}}{P_{\rm b}}\right)^{\rm GW}\!\!\!\!\!\!\!
    -\left(\frac{\dot D}{D}\right)
    \end{equation}
where we have ignored possible contributions due to mass loss or tidal
effects in the system \citep[see][]{lk05}.

\begin{figure}[htb]
\centering
\includegraphics[width=0.5\textwidth]{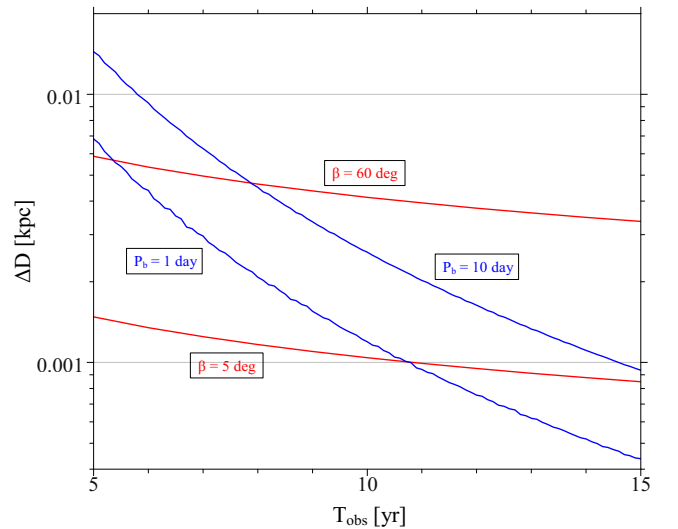} 
\caption{Precision for distance measurements inferred from the timing
  of a pulsar at 1 kpc distance as a function of observing time
  span. The red curve is for the ``classical'' timing parallax at two
  different ecliptic latitudes, $\beta$. The distance uncertainties
  achieved with the kinematic parallax are shown in blue for two
  different orbital periods.  We again assume weekly observations
  with a TOA precision of 10 ns.
\label{fig:wex6}}
\end{figure}

The term $\dot D/D$ can be determined from Eq.~\ref{equ:PBDOBS} by
comparing the computed GR value with the
observations. Figure~\ref{fig:wex4a} compares the relative size of the
contributing terms by showing the orbital period change of a pulsar in
a 1-d orbit with a $0.3M_\odot$ companion and a typical transverse
velocity of 100 km s$^{-1}$. The red curve shows the kinetic
contribution of the Shklovskii term, the green one representing the
gravitational wave damping term and the blue line compares the timing
precision of three different time spans of observations, namely 5, 10,
and 15 years, respectively. Figure~\ref{fig:wex6} compares the
resulting absolute uncertainty in distance measurement for the
``classical'' timing parallax (red) at two different ecliptic
latitudes with the results for the ``kinematic'' parallax (blue) with
different orbital periods. It is interesting to see that for the first
few years of observations the classical timing parallax yields much
higher precision but that in the long-term the kinematic parallax
eventually leads to more precise results assuming that the
acceleration in the Galactic gravitational potential can be accounted
for and that the masses are known with sufficient accuracy via the
measurement of PK parameters.

\section{Results}
\label{sec:results}

While we have demonstrated the SKA capabilities in principle, we now
review the applicability of our results to real SKA observations.

\begin{figure}[htb]
\centering
\includegraphics[width=0.35\textwidth, angle=-90]{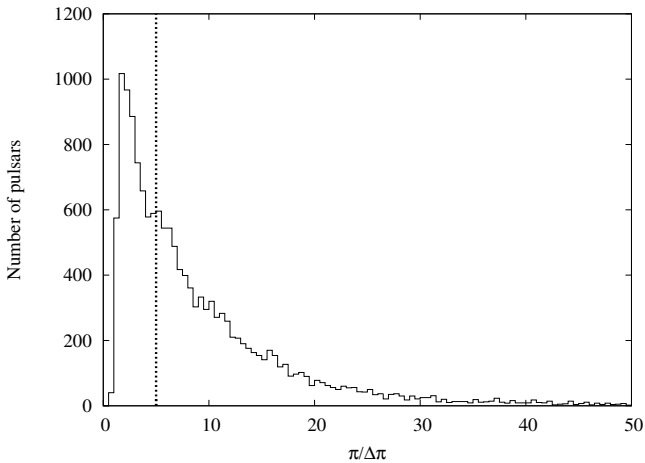}
\caption{Histogram of $\pi/\Delta\pi$ for the pulsars detected in the
  simulation of imaging observations. The vertical dotted line marks
  the $\pi/\Delta\pi = 5$ cutoff.}
\label{fig:parallax1}
\end{figure}

\subsection{Imaging observations}

Fig.~\ref{fig:parallax1} shows the histogram of the quantity
$\pi/\Delta\pi$, where $\pi$ is the parallax and $\Delta\pi$ is the
estimated error in the parallax, for the simulated sample. The
histogram corresponds to two underlying distributions for which the
parallax is limited by the SNR of the pulsar or the dynamic range. The
low number of pulsars around $\pi/\Delta\pi\sim 1$ arises from the
detection limit of the SKA survey. The main peak corresponds to the
large number of pulsars with a low SNR. The rise in the distribution
at $\pi/\Delta\pi\sim 5$ reflects the peak of the pulsar distance
distribution at $\sim$6\,kpc. Taking a cutoff $\pi/\Delta\pi = 5$, we
find that the SKA can potentially measure the parallaxes for
$\sim9,000$ pulsars with an error of 20\% or smaller. This includes
pulsars out to a distance of 13\,kpc. For the Galactic plane, as many
as 25 pulsars can be observed in the FoV when using a single-pixel receiver. If
we assume that the correlator can deal with all the pulsars in the
FoV, it will take 215 days to measure the parallaxes of all the
pulsars (9\,000) in the Galactic plane and 450 days to measure the
parallaxes of all the detectable pulsars to a parallax error of 20\%.

\begin{figure}[htb]
\centering
\includegraphics[width=0.5\textwidth]{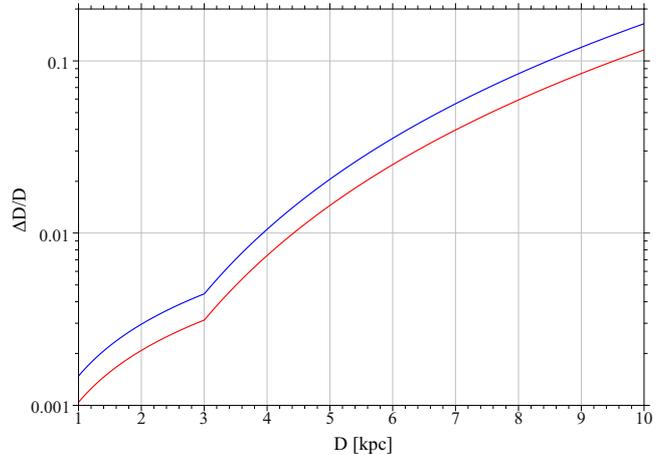}
\caption{ Fractional precision of distance measurement by timing
  parallax for weekly observations.  The TOA error is assumed to be
  10\,ns up to 3\,kpc beyond which the error scales with distance
  squared.  The blue line indicates five years of observations, the red
  one ten years, respectively.}
\label{fig:wex3b}
\end{figure}

\subsection{Timing observations}

We have shown that timing observations of pulsars can yield a parallax
precision that exceeds that of an imaging parallax by at least an
order of magnitude in terms of both relative and absolute
precision. For longer time spans, the kinematic parallax is of even
higher accuracy. For these results, we have typically assumed a TOA
uncertainty of 10 ns. Such a precision is only possible for recycled
millisecond pulsars (MSPs) and is a factor of 5--10 higher than what
has been achieved for the best MSPs to date. Increased sensitivity is
expected to help improve the timing precision in a linear way,
providing an up to 100 times higher sensitivity with the SKA and
ensuring that such timing precision appears easily achievable
\citep{lk05}. Uncertain factors are, however, possible intrinsic
timing noise of the pulsar clock, propagation effects in variable
interstellar weather, and both instrumental and improper calibration
effects. Recent and ongoing studies \citep[][Kuo et al. in
  prep.]{Verbiest09} indicate that despite challenges, a timing
precision for MSPs between 10 and 50 ns seems feasible. We expect that
about $\sim 6000$ of the pulsars detected with the SKA will be MSPs
\citep{Smits09} but not all of those will be suitable for high
precision timing. Moreover, as the new pulsars become more distant,
the flux density will decrease and hence the timing precision becomes
worse. We attempt to take this into account by increasing the TOA
errors by a factor of $(d/d_0)^2$, where $d_0$ determines the distance
beyond which the timing precision is dominated by the radio meter
equation (rather than other limiting factors). Inspecting the timing
precision of the presently known MSPs as a function of known distance,
we estimate this transition to be around 3 kpc. This number may be
significantly larger for the SKA, so that the present results using
classical timing parallax shown in Fig. \ref{fig:wex3b} are probably a
conservative estimate. These suggest that distances can be measured
with a precision of at least 10\% to a distance of about 10 kpc.

\begin{figure}[htb]
\centering
\includegraphics[width=0.5\textwidth]{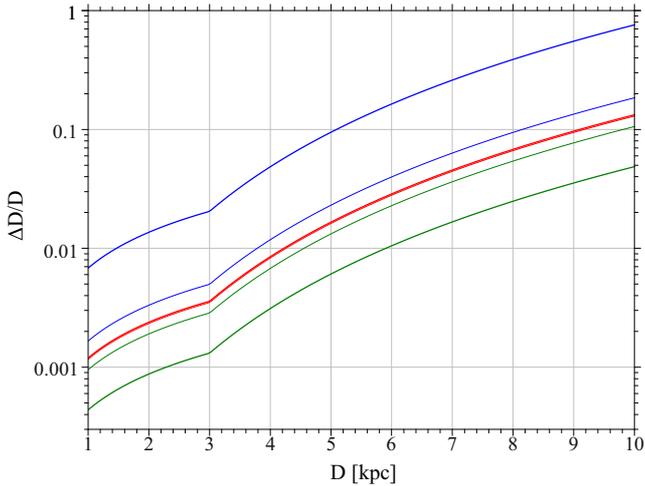}
\caption{Fractional precision of a distance measurement for weekly
  observations with TOA errors of 10 ns up to a distance of 3
  kpc. Beyond 3 kpc, we assume the timing precision to become worse by
  a factor scaling with the square of the distance. Colours indicate
  the length of the observing span: 5 yr (blue), 10 yr (red), and 15 yr
  (green), respectively. The thick lines indicate the kinematic
  distance measurement for a pulsar in a 1 day orbit with an 0.3 solar
  mass companion. The thin lines indicate the timing parallax for the
  same pulsar. Note that the 10-yr lines (red) overlap.}
\label{fig:wex5b}
\end{figure}

In Fig. \ref{fig:wex5b}, we compare the distance measurements using
classical timing parallax and using kinematic parallax and adjusting
the TOA precision for the distance scaling. As before, the classical
timing parallax yields the superior results for the first years of
observations, but is overtaken by the kinematic parallax after 10
years.

To gauge how many MSPs would be likely to reach the timing precision
assumed above, we compare the obtained precision among a consistent
group of known MSPs at a relatively well-determined distance by
inspecting the timing results for those 16 MSPs in the globular
cluster 47 Tucanae that are regularly observed. The other eight MSPs
in the cluster are at the current limit of detectability of the Parkes
telescope and can only be timed when the interstellar scintillation
boosts the flux density above the detection threshold
\citep{Freire03}.  Table~\ref{tab:MSP} shows the current and predicted
TOA uncertainty as well as an estimate of the accuracy with which the
parallax of these 16 MSPs can be measured with the SKA.  These
estimates are based on a scaling that ensures that the TOAs are
improved by a factor of $\sim100$.  However, as the full SKA may not
be able to be phased up for timing observations, we conservatively
estimate that we can only use the inner 5 km, containing 50\% of the
collecting area, so that we scale the TOA errors by a factor of 50. To
estimate the parallax error inferred from timing each of the 6\,000
new MSPs that will be found by the SKA, we randomly assign a TOA error
from the 16 MSPs in 47 Tucanae. The parallax error is then corrected
for ecliptic latitude and distance following the results shown in
Figs.~\ref{fig:wex1a} and ~\ref{fig:wex5b}.

\begin{figure}[htb]
\centering
\includegraphics[width=0.35\textwidth, angle=-90]{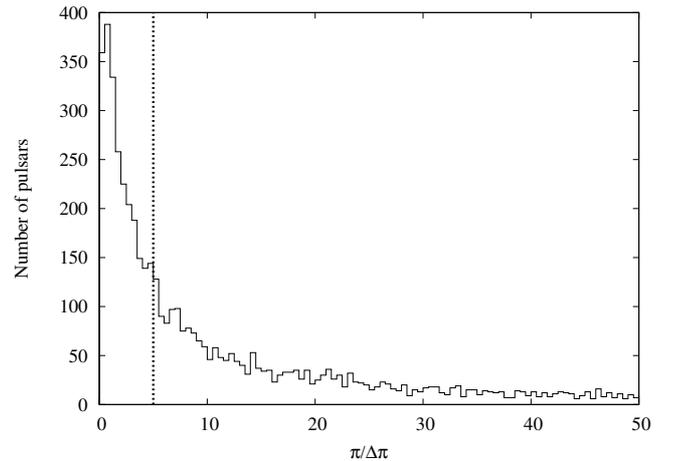}
\caption{Histogram of $\pi/\Delta\pi$ for the pulsars detected in the
  simulation of timing 6\,000 MSPs. The vertical dotted line indicates
  the $\pi/\Delta\pi = 5$ cutoff.}
\label{fig:parerr_msp}
\end{figure}

Figure~\ref{fig:parerr_msp} shows the histogram for the quantity
$\pi/\Delta\pi$, where $\pi$ is the parallax determined by means of
timing MSPs and $\Delta\pi$ is the estimated error in the parallax for
the simulated sample. Taking once again a cutoff $\pi/\Delta\pi = 5$,
we find that the SKA can potentially measure the parallaxes for
$\sim3\,600$ MSPs with an error of 20\% or smaller out to a distance
of 9\,kpc. Timing all these MSPs using the phased array feeds of the
SKA with an observation time of 1 hour, will take roughly 25 days. To
obtain a timing solution requires re-observations of these MSPs once
every two weeks over a period of at least one year. The total
observation time would then amount to 650 days. If we were to time
these MSPs only to achieve a parallax error of 20\%, the total
observation time would drop to roughly 40 days.  In any case, there
will be a follow-up timing for each MSP that the SKA will detect to
select the most interesting MSPs. Therefore, obtaining parallaxes for
many of these MSPs will not require extra observation time.

\begin{table}
  \caption{Error in TOAs from 16 MSPs from the cluster 47 Tucanae
    and the resulting error in parallax from the timing solution.}
  \begin{tabular}{llll}
    \hline
    \hline
    \scr{MSP} & \scr{Error in TOA}    & \scr{Estimated error}   & \scr{Error in} \\
           & \scr{as measured}        & \scr{in TOA when}       & \scr{parallax}\\
            & \scr{by Parkes ($\mu$s)} & \scr{measured with}    & \scr{($\mu$as)}\\
           & \scr{(Freire et al.}   & \scr{the SKA ($\mu$s)}    & \\
           & \scr{2003)}            &                           &  \\ 
    \hline
    J0024$-$7204C  & 11 & 0.22  & 38\\ 
    J0024$-$7204D  &  6 & 0.12  & 21\\
    J0024$-$7204E  &  9 & 0.18  & 31\\
    J0024$-$7204F  & 10 & 0.20  & 34\\
    J0024$-$7204G  &  7 & 0.14  & 24\\
    J0024$-$7204H  & 20 & 0.40  & 69\\
    J0024$-$7204I  & 14 & 0.28  & 88\\
    J0024$-$7204J  &  3 & 0.06  & 10 \\
    J0024$-$7204L  & 32 & 0.64  & 110\\
    J0024$-$7204M  & 17 & 0.34  & 58\\
    J0024$-$7204N  & 12 & 0.24  & 41\\
    J0024$-$7204O  &  8 & 0.16  & 28\\
    J0024$-$7204Q  & 19 & 0.38  & 65\\
    J0024$-$7204S  & 11 & 0.22  & 38\\
    J0024$-$7204T  & 49 & 0.98  & 168\\
    J0024$-$7204U  &  9 & 0.18  & 31\\
    \hline
  \end{tabular}
  \label{tab:MSP}
\end{table}

\subsection{Pulsar proper motions}
Astrometry with the SKA will permit us to obtain accurate measurements
of the proper motion of pulsars. These are extremely valuable for
constraining the origin of pulsar velocities, which is currently known
with an accuracy of only a few to tens of mas per year for normal
pulsars and often $<$ 1 mas per year for MSPs \citep{lk05}. Using
imaging observations with the SKA allows a potential accuracy of
15\,$\mu$as per year. From our simulations, we find that for 60\% of
all pulsars the proper motion can be measured with an accuracy of
$15-25$\,$\mu$as per year. Furthermore, we find that using timing
observations, the SKA can achieve a proper motion accuracy of $<
5\mu$as per year for almost 1\,000 MSPs and an accuracy of $< 30\mu$as
per year for more than 3\,000 MSPs.

\section{Discussion}

\label{sec:discussion}
We have investigated two methods to measure the parallax of radio
pulsars using the SKA. The imaging parallax method utilises the long
baselines of the SKA to keep track of the position of the pulsar on
the sky during different orbital positions of the Earth. The timing
method uses the arrival times of the radio signals of pulsars to
derive a ``classical'' parallax, an ``orbital-annual'' parallax, or a
``kinematic'' parallax. With imaging parallax, it is possible to
determine parallaxes with an error of 20\% or smaller for 9\,000
pulsars out to a distance of 13\,kpc. Measuring the parallaxes of all
the pulsars in the Galactic plane alone will require 215 full days of
observing. Furthermore, to obtain the required astrometric accuracy of
15\,$\mu$as requires several calibrators to be within a few
arc-minutes of the source.  With the timing parallax method,
parallaxes can be measured out to 9\,kpc for about 3\,600 MSPs, with
20\% or higher accuracy. This would require a minimum of roughly 40
days of observation time. However, there will have to be follow-up
timing in any case for each MSP that will be detected by the SKA to
select the most interesting MSPs. Therefore, obtaining parallaxes for
many of these MSPs will not require extra observation time.

When we consider the increased timing precision with the SKA and the
recent realization that the apparent ``timing noise'' in
non-millisecond pulsars is non-random and of magnetospheric origin,
possibly allowing to correct for it \citep{lhk+10}, it seems likely
that parallax measurements will even be possible for normal,
non-recycled pulsars. In this case, the sample of available pulsars
could easily exceed the number of 10\,000 but this is difficult to
estimate at this stage.

The imaging and timing parallax methods are complementary, with timing
parallax providing high quality measurements for a relatively small
number of objects but to large distances, whereas imaging parallax can
provide good results for a large number of closer objects.  Timing
parallax is limited to millisecond pulsars with stable timing
characteristics, whereas an imaging parallax depends only on the
observed strength of the radio emission, not the rotation
characteristics of the pulsar. Imaging parallax measurements can
indeed help us significantly with precision tests of general
relativity to correct for the same kinematic effects that we can
otherwise use for precise distance measurements. Moreover, obtaining
imaging information immediately after the discovery of a new pulsar
will help us to obtain positional information that will help the
typical ``solving procedure'' of a pulsar, i.e.~the first
determination of a coherent timing solution.  In general, solving
pulsars takes several weeks to months of intensive observing, but the
positional information from imaging eradicates the correlations
between position and period derivative and allows us to reduce
the cadence of the observations significantly.

From the above results, it is clear that the SKA will become a superb
astrometry instrument, the data of which feeds directly back into
fundamental astrophysical questions. For instance, the distance to
about 9\,000 pulsars can be measured with an error of 20\% or
smaller. The dispersion measure from each pulsar can then be used to
accurately calculate the electron density between the Earth and each
of these pulsars. This will result in a map of the distribution of the
ionised gas of high accuracy out to a distance of 13\,kpc. By timing
the most stable MSPs that the SKA will discover, the ionised gas can
be mapped to a distance of 9\,kpc.  By imaging pulsars, we can locate
them with a precision that is comparable with that of GAIA, which is
able to measure distances out to 10 kpc with a precision of 10\%.
This is also achievable by timing MSPs and may be able to be, for
selected sources, improved upon.  This is exciting as we will be able
to get independent distance measurements from GAIA and the SKA for
pulsars that have an optical companion or are located, e.g.  in a
globular cluster.

The simulations presented in this paper are not corrected for the
Lutz-Kelker bias \citep{Lutz93}. This bias is introduced when the
parallax measurement does not take into account the larger volume of
space that is sampled at smaller parallax values and leads to a
systematic overestimate of the parallax, thus to an underestimate of
the distance to the star. This bias is strongly dependent on the
accuracy of the parallax measurement, becoming smaller as the
measurement becomes more accurate. ~\citet{vlm10} have studied this
bias for the specific case of the parallax measurements of neutron
stars, incorporating the bias introduced by the intrinsic radio
luminosity function and a realistic Galactic population model for
neutron stars. They found a significant bias for measurements with a
50\% error. In the present paper, we consider parallax accuracies of
20\% and higher. We find that in our simulations the Lutz-Kelker bias
is always smaller than one standard deviation and can be neglected for
most pulsars, hence does not affect our conclusions. Nevertheless,
since the bias in a pulsar parallax measurement depends on the
luminosity of the pulsar and the position on the sky, it is advisable
to correct for it in every pulsar parallax measurement\footnote{The
  Lutz-Kelker bias for pulsars can be determined online:
  \url{http://psrpop.phys.wvu.edu/LKbias}}.

Overall, our result shows that the SKA requires both a concentration of
collecting area in the core, to search for and detect the Galactic
pulsar population and then measure accurate timing parallaxes, and a
distribution of collecting area on long baselines, to achieve
the angular resolution required to perform imaging parallax
measurements.  Thus, the current design specifications for the SKA are
adequate for achieving both of these objectives.

\acknowledgement{The authors would like to thank an anonymous referee
  for his/her comments. This effort/activity is supported by the
  European Community Framework Programme 6, Square Kilometre Array
  Design Studies (SKADS), contract no 011938.  We gratefully
  acknowledge support from ERC Advanced Grant ``LEAP'', Grant
  Agreement Number 227947 (PI Michael Kramer). The International
  Centre for Radio Astronomy Research is a Joint Venture between
  Curtin University and The University of Western Australia, funded by
  the Western Australian State Government.}

\bibliographystyle{aa} 
\bibliography{parallax}

\end{document}